\documentclass[reprint,amsmath,amssymb,aps,pra,groupedaddress,showpacs]{revtex4-1}
\usepackage{graphicx}%
\usepackage{dcolumn}%
\usepackage{bm}%
\usepackage[
  letterpaper, mag=1000,
  left=2.5cm, right=1cm, top=2cm, bottom=2cm, headsep=0.7cm, footskip=1cm
]{geometry}

\begin{document}

\title{Unidirectional Amplification and Shaping of Optical Pulses by Three-Wave Mixing with Negative Phonons} 

\author{Alexander K. Popov$^{1}$}%
\email{apopov@uwsp.edu}
\homepage{http://www4.uwsp.edu/physastr/apopov/}%
\author{
Mikhail I. Shalaev$^{2}$
}%
\author{
Sergey A. Myslivets$^{3}$ }%
\author{
Vitaly V. Slabko$^{2}$}%
\affiliation{$^1$University of Wisconsin-Stevens Point, Stevens Point WI 54481, USA  \\
and Birck Nanotechnology Center,
Purdue University,
West Lafayette, IN 47907, USA\\
$^2$Siberian Federal University,
660041 Krasnoyarsk, Russian Federation\\
$^3$Institute of Physics,
Siberian Branch of the
Russian Academy of Sciences,  660036 Krasnoyarsk, Russian Federation}%
\begin{abstract}
A possibility to greatly enhance frequency-conversion efficiency of stimulated Raman scattering is shown by making use of extraordinary properties of three-wave mixing of ordinary and backward waves. Such processes are commonly attributed to  negative-index plasmonic metamaterials.  This work demonstrates the possibility to replace such metamaterials that are very challenging to engineer by  readily available crystals which support elastic waves with contra-directed phase and group velocities. The main goal of this work is to investigate specific properties of  indicated nonlinear optical  process in short pulse regime and to show that it enables elimination of fundamental detrimental effect of fast damping of optical phonons on the process concerned. Among the applications  is the possibility of creation of a family of unique photonic devices such as unidirectional Raman amplifiers and femtosecond pulse  shapers with greatly improved operational properties.
\end{abstract}

\maketitle

\section{Introduction}

Extraordinary features of coherent nonlinear optical NLO energy conversion processes in negative-index metamaterials (NIMs) that stem from wave-mixing of ordinary and backward electromagnetic waves (BEMW)  and the possibilities to apply them for compensating optical losses  have been investigated in \cite{APB,OL,OLM,APL,APB09,OL09,SPI,JOA,WAS,ch}. Essentially different properties of three-wave mixing (TWM)  and four-wave mixing processes on one hand and second harmonic  and third harmonic generation  have been revealed in \cite{EPJD,SPIE,KivSHG,Sc,SHG,APB,Kaz,EPJST,El}. Ultimately, it was shown that  NLO propagation processes that involve BW enable a great enhancement of energy-conversion rate at otherwise  equal nonlinearities and intensities of input waves. A review can be found in \cite{EPJD,SPIE}.

In \cite{ph}, three-wave mixing (TWM) of continuous waves (CW) was investigated in the scheme of stimulated Raman scattering (SRS) as an analog  of parametric interaction of waves in a medium with negative dispersion. Two of the coupled waves were ordinary electromagnetic waves and the third one was backward elastic wave corresponding to optical phonons which exhibit  negative dispersion $d\omega_v(k)/dk$. The latter gives rise to contra-directed group and phase velocities of propagation of lattice vibrations, i.e. backward elastic waves. Such waves exist in crystals  containing more than one atom per unit cell.  Indeed, backwardness is a main property of EM  waves propagating in NIMs. Such BWs were predicted by L. I. Mandelstam in 1945 \cite{Ma}, who also had pointed out that negative refraction is a general property of the BWs. Optical phonons with wave vector much smaller than the reciprocal lattice vector have been considered in \cite{ph}. The focus  was placed on a coupling scheme where the fundamental radiation was converted to co-propagating Stokes and contra-propagating vibrational waves under the phase-matched conditions. The possibility of the extraordinary SRS  was shown, which is typical for TWM whereas one of the coupled waves possesses negative dispersion \cite{APB,OL}. The latter results in  gain of the Stokes component along the medium much greater than the exponential growth.

The basic idea underlying the proposed concept is as follows.
\begin{figure}[h]
\begin{center}
\includegraphics[width=.8\columnwidth]{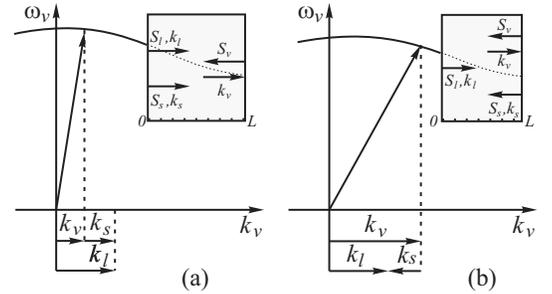}
\caption{\label{phf1}  Negative dispersion of optical phonons and two phase matching options for short- and long-wave vibrations: (a) -- co-propagating, (b) -- contra-propagating  fundamental, control,  and Stokes, signal, waves. Insets: relative directions of the energy flows and the wave-vectors. } \end{center}
\end{figure}
The dispersion curve $\omega(k)$ for optical phonons is depicted in Fig.~\ref{phf1}. It is negative  in the range from zero to the boundary of the first Brillouin's zone. Hence, the group velocity of such phonons, $\mathbf{v}_{v}^{gr}$, is antiparallel with respect to its wave-vector, $\mathbf{k}_{v}^{ph}$, and phase velocity, $\mathbf{v}_{v}^{ph}$, because
\begin{equation}
{\mathbf{S}}={\mathbf{v}_g}U,\quad {\mathbf{v}_g}=({\mathbf{k}}/k)[\partial\omega/\partial k],\quad \partial \omega(k)/\partial k < 0. \label{gr}
\end{equation}
Optical vibrations can be excited by the light waves due to the two-photon (Raman) scattering. The latter gives the ground to consider such a crystal as the analog of the medium with negative refractive index at the phonon frequency and to employ the processes of parametric interaction of three waves, two of which are ordinary EM  waves and the third is the wave of elastic vibrations with the directions of the energy flow and of the wave-vector opposite to each other. For continuous wave SRS in the coupling geometry depicted in Fig. \ref{phf1}(a) and diamond crystal,  estimates made in \cite{ph} have shown that the minimum intensity of the fundamental field $I_{\min}$, required to reach the aforementioned extraordinary Raman amplification is on the order of $I_{\min}\sim 10^{18}$~W/cm$^2$, which exceeds  optical breakdown threshold.  The main process that determines such a high intensity is fast phonon damping, which is characterized by the optical phonon relaxation rate on the order of  $\tau_v
=6\cdot 10^{-12}$~s, and relatively small group velocity of the elastic
oscillation $v_v \sim 10^2\div10^3$ cm/s for optical phonons with small optical wave vectors \cite{An,Chen}. Diamond is a representative of  a wide class of transparent crystals which support optical phonons \cite{An,Chen,Card,Alf,Gor}.

The \emph{goal of  this work} is to show that this seemingly formidable obstacle can be removed and the threshold intensity of the fundamental field $I_{\min}$  required to realize such extraordinary coupling can be significantly reduced. It is possible by making use of short laser pulses with duration $\tau_p$ less then  the lifetime of the vibrational oscillation of $\tau_v$. Some of the properties of the output fundamental and Stokes pulses, such as duration and pulse shape are investigated.
\section{Equations and model}
Consider  lowest-order SRS process \cite{ShB,Boy}. The  waves are given by equations
\begin{eqnarray} \label{eq1}
 E_{l,s} &=&({1}/{2})\varepsilon _{l,s} (z,t)e^{ik_{l,s} z-i\omega _{l,s}
t}+c.c. ,\nonumber\\
 Q_v &=&({1}/{2})Q(z,t)e^{ik_v z-i\omega _v t}+c.c.
\end{eqnarray}
Here, $\varepsilon _{l,s} $,  $Q$, $\omega _{l,s,v} $ and $k_{l,s,v} $ are the amplitudes, frequencies and wave-vectors of the  fundamental, Stokes and vibrational waves;  $Q_v (z,t)=\sqrt \rho x(z,t)$; $x$ is displacement of the vibrating particles,  $\rho $ is the medium density. With account for the frequency and phase matching exppressions,
\begin{equation*} 
 \omega _l =\omega _s +\omega _v \left( {k_v } \right) ,\quad
 \vec {k}_l =\vec {k}_s \left( {\omega _s } \right)+\vec {k}_v,
\end{equation*}
one obtains  following partial differential equations for the slowly varying amplitudes in the approximation of the of first order of $Q$ in the polarization expansion \cite{Bl}:
\begin{eqnarray}
\label{eq2}
 \frac{\partial \mathcal{E} _l }{\partial z}+\frac{1}{v_l}\frac{\partial \mathcal{E} _l }{\partial t}&=&i\frac{\pi \omega _l^2 }{k_l
c^2}N\frac{\partial \alpha }{\partial Q}\mathcal{E} _s Q \nonumber\\
 \frac{\partial \mathcal{E} _s }{\partial z}+\frac{1}{v_s
}\frac{\partial \mathcal{E} _s }{\partial t}&=&i\frac{\pi \omega _s^2 }{k_s
c^2}N\frac{\partial \alpha }{\partial Q}\mathcal{E} _l Q\ast \nonumber\\
 \frac{\partial Q}{\partial z}+\frac{1}{v_v }\frac{\partial Q}{\partial
t}+\frac{Q}{l_v}&=&i\frac{1}{4\omega _v v_v }N\frac{\partial
\alpha }{\partial Q}\mathcal{E} _l \mathcal{E} _s^\ast.
\end{eqnarray}
Here, $v_{l,s,v} $ are the projections of the group velocities of the fundamental, Stokes and vibration waves on the z-axis,  $N$ is the number density of the vibrating molecules, $\alpha $ is the molecule polarizability,  $l_v=\tau_v v_v$ is the mean free path of phonons.

For the sake of simplicity, we consider model of a rectangular pulse of input fundamental radiation with duration much shorter that the phonon lifetime $\tau_v$.
In the  coordinate frame associated with this pulse and within the fundamental pulse range, complex amplitudes of  two other
interacting fields become time independent, and Eqs.~(\ref{eq2}) transform into set of  ordinary differential equations whose solution is known in the approximation of constant pump amplitude or can be relatively easily found numerically when depletion of the pump is accounted for. Here, the boundary conditions must be fulfilled not at the boundaries of the medium but at the
boundaries of the fundamental pulse. The latter is correct for the period of time after the instant when generated waves reach  the boundaries of the fundamental pulse due to the difference in their group velocities or direction of propagation. Such approximation becomes true after travailing  a distance $l>L_g^{\max } $, where
$L_g^{\max } =\max \{L_g^s ,L_g^v \}$. Here, $L_g^{s,v}$, which is  further referred to as group length, is defined as
\begin{equation}\label{eq3a}
L_g ={\tau_p v_l^2}/{\left| v_{s,v} - v_l \right|},
\end{equation}
where   $v_{s,v}$ are
negative if their energy fluxes are opposite to that of the   fundamental wave.
Hereinafter, the waves are referred to  as co-propagating waves if Poynting vector of the Stokes wave is co-directed with that of fundamental wave (Fig.~\ref{phf1}a), and as counter-propagating in the opposite case (Fig.~\ref{phf1}b).
In the approximation of constant pump amplitude, in the coordinate frame  locked to the pump pulse and within the pulse, equations for generated Stokes and backward vibration waves take the form:
\begin{eqnarray}
\label{eq3}
 d\,Q/d\xi &=&-ig_v \mathcal{E} _s^* -Q K_v/l_v \nonumber\\
d\,\mathcal{E}_s /d\xi&=&ig_s Q^*,
\end{eqnarray}
where  $g_v=-K_v N(d\alpha/dQ)\mathcal{E} _l /( 4\omega _v v_v)$, $K_{v,s}=v_{v,s}/(v_{v,s}-v_l )$,\quad
$g_s=K_s N(d\alpha/dQ)\mathcal{E}_l \pi \omega_s^2/( k_s c^2)$.
Here, intensity of the laser beam is taken constant. Depletion of the fundamental beam is neglected. Equations (\ref{eq3}) may describe significant amplification of the  Stokes signal, however they remain valid until only relatively small part of the strong input laser beam is converted.
Equations (\ref{eq3}) do not depend on time and are
similar to those describing  CW TWM  \cite{ph}.
Here, group velocities $v_v<0$, $v_s>0$ and wave vector $k_s>0$  for co-propagating and $v_s<0$, $k_s<0$ for counter-propagating phase matching configurations.  Since the Stokes frequency is less than that of the fundamental one,  $v_s>v_l$  and is much greater than the magnitude of phonon group velocity $v_v$.In the case of co-directed laser and Stokes waves, the boundary conditions  take the form:
\begin{eqnarray}
\label{eq4}
\mathcal{E} _s (\xi =0)=\mathcal{E} _s^0, \quad 
 Q(\xi =l_p )=0.
\end{eqnarray}
In the opposite case of counter-propagating Stokes and laser waves, the boundary conditions are given by equations
\begin{eqnarray}
\label{eq5}
 \mathcal{E} _s (\xi =l_p )=\mathcal{E} _s^{l_p }, \quad
 Q(\xi =l_p )=0.
\end{eqnarray}
Here, $l_p =\tau _p v_l $ is length of the fundamental pulse.

Solution to  Eqs.~(\ref{eq3}) can be presented in the form:
\begin{eqnarray}
\label{eq7}
&T_s^{\uparrow \uparrow }&=\left| \frac{e^{\gamma' \xi'}\left\{ R_1 \cos Z+\gamma' \sin Z \right\}}
{R_1 \cos \left( {R_1}\right)+\gamma'\sin \left( R_1 \right)} \right|^2, \nonumber\\
&T_s^{\uparrow \downarrow }&=\left| \frac{ \beta _1 e^{\beta _2( 1-\xi')}-
 \beta _2 e^{\beta _1 \left(1-\xi '\right)}} {2R_2}  \right|^2.
\end{eqnarray}
Here,
$ T_s^{\uparrow \uparrow } =\left| {{\mathcal{E} _s (\xi
)}/{\mathcal{E} _s^0 }} \right|^2$ and $T_s^{\uparrow \downarrow } =\left| {{\mathcal{E} _s (\xi )}/{\mathcal{E}
_s^{l_p } }} \right|^2$ are transparency (amplification) factors,
 $\gamma'=\gamma l_p$, $\gamma={- K_v }/( 2l_v)$,  $R_{1,2} =\sqrt {g'^2\mp \gamma'^2}$, $Z=R_1(1-\xi')$, $\xi'=\xi/l_p $,  $\beta_{1,2} =\gamma\pm R_2$,
$g'=gl_p$, $g=\sqrt {g_v^* g_s }$.

For $\gamma \ll g$,  phonon damping can be
neglected and Eqs.~(\ref{eq7}) take the form
\begin{eqnarray}
\label{eq8}
 T_s^{\uparrow \uparrow }&=&\left| \frac{\cos \left[ {gl_p(1-\xi ')}
\right]}{\cos \left( {gl_p } \right)} \right|^2 ,\nonumber\\
 T_s^{\uparrow \downarrow}&=&\left| \left[ e^{gl_p(1-\xi')}+e^{-gl_p( 1-\xi ')} \right]/2 \right|^2 .
\end{eqnarray}

\section{Enhancing coherent energy transfer between electromagnetic waves through backward optical phonons}

In the given approximation of neglected depletion of the fundamental wave,  solution to  Eq.~(\ref{eq8}) tends to infinity for
certain  pulse length and intensity. This indicates the possibility to greatly enhance the conversion efficiency provided that the following requirements are met:
\begin{equation}
\label{eq9}
\cos \left( {gl_p} \right)=0;\quad l_p = ( \pi/2)/\sqrt{g_v^*g_s}
\end{equation}
Corresponding threshold intensity of the fundamental field $I_{\min }^p$, which is required for realization of great enhancement of energy conversion due to NLO coupling with BW phonon,  is given by equations:
\begin{equation}
\label{eq11}
g>\gamma,\quad I_{\min }^p =\frac{K_v}{K_s }\frac {cn_s \lambda _{s0} \omega _v}
{16\pi ^3v_v \tau_v ^2}\left| N\frac{\partial \alpha}{\partial Q}\right|^{-2}
\end{equation}
Here, $n_s$ is refractive index at Stokes frequency, $\lambda_{s0}$ is Stokes wavelength in the vacuum.

\subsection{Factors discriminating BW SRS in continuous wave and pulsed regimes}
Threshold intensity of fundamental radiation required for realization of BW SRS in CW regime is given by \cite{ph}
\begin{equation}
I_{\min } =\left({cn_s\lambda_{s0}\omega_v}/{8\pi^3l_p\tau }\right)\left| {N\partial \alpha /\partial Q}  \right|^{-2}.
\end{equation}
It differs from the threshold value in  pulse regime by factor
\begin{equation}
\label{eq12}
\frac{I_{\min }^p }{I_{\min } }=-\frac{K_v}{K_s }\approx -\frac{v_v }{v_l }\frac{v_s -v_l }{v_s }.
\end{equation}
For the same typical crystal parameters as employed in \cite{ph},  Eq.~(\ref{eq12}) yields $I_{\min }^p /I_{\min } \approx 10^{-11}$.
 Hence, $I_{\min }^p $ decreases down to $I_{\min }^p \sim 10^7$~W/cm$^2$, which is achievable with commercial femtosecond lasers and falls below optical breakdown threshold for most transparent crystals.

Equation~(\ref{eq12}) displays  two factors that determine
substantial decrease of $I_{\min}^p$ in pulsed regime compared to that in CW one. First factor is the ratio of group velocity of the elastic wave to that of the fundamental one, $v_v/v_l $, which is on the order of $\sim 10^{-8}$. This factor is attributed to the fact that phonons generated on the front edge of the laser pulse propagate in the opposite direction and, hence, exit very fast, practically with the optical group velocity, from the fundamental pulse zone before it is dissipated.  Hence, effective phonon mean free pass becomes commensurable with the fundamental pulse length.  This mitigates the detrimental effect of phonon damping. With increase of $v_v$, phonon mean free path grows, which decreases both  $I_{\min }^p$ and  $I_{\min }$ in a way that the advantage of pulse regime over CW regime diminishes.
The second factor in Eq.~(\ref{eq12}) determines  further decrease of $I_{\min }^p $  due to small optical dispersion in the transparency region of the crystals. The fact that the Stokes pulse surpasses  the fundamental one slowly  increases   significantly the effective NLO coupling length.

\subsection{Numerical simulations, $\tau_p\ll\tau_v$.}\label{sec:calc}

Numerical analysis is done for the model with  parameters typical for diamond crystal \cite{An,Chen}: carrying wavelength of the fundamental pulse $\lambda=800$~nm, pulse duration  $\tau_p=60$~fs, $\omega_v= 1332$~cm$^{-1}$,  vibrational
transition width $(c\tau_v)^{-1}= 1.56$~cm$^{-1}$, $v_l=1.228\cdot10^{10}$~cm/s, $v_s=1.234\cdot10^{10}$~cm/s, $v_v=100$~cm/s for co-propagating and  $v_v=2000$~cm/s for counter-propagating waves, $Nd\alpha/dQ=3.78\cdot10^7$~(g/cm)$^{1/2}$.
Partial differential equations (\ref{eq2}) were solved numerically in three steps: TWM in the vicinity of the entrance to the  Raman medium, TWM and propagation through the medium  and TWM in the vicinity of the exit from the Raman slab. Numerical simulation for the first and third medium intervals were made in the laboratory reference frame with the boundary conditions applied to the corresponding edges of the slab.  NLO propagation process inside the slab was simulated in the moving frame of reference  with the boundary conditions applied to the pulse edges. Such a computing approach  allows for significant reduction of the computation time since, for each given instant, integration is required only through a space interval covered by the fundamental pulse and not through the entire medium.
 Shape of the fundamental pulse was chosen nearly rectangular and symmetric with respect to its center
$$\mathcal{E}_l =\frac{1}{2}\mathcal{E}_l^0 \{\tanh[( t_0+t_p-t)/t_f]-\tanh[(t_0-t)/t_f]\},$$
the slope  $t_f =0.1$, the pulse duration at half-maximum $t_p=1$ and pulse delay $t_0 =0.6$ were scaled  to the fundamental pulse width $\tau_p$. The amplitude of input CW Stokes signal was chosen $\mathcal{E}_s^0=10^{-5}\mathcal{E}_l^0$.
\begin{figure}[t]
\begin{center}
\includegraphics[width=.8\columnwidth]{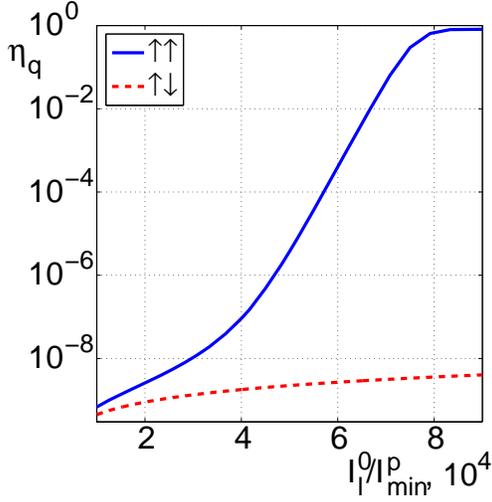}
\caption{\label{f5} Quantum conversion efficiency vs. intensity (energy) of the input  pulses for co-propagating and contra-propagating phase matching geometries.}
\end{center}
\end{figure}

Figure \ref{f5} displays  output quantum conversion efficiency
 $\eta_{sq}=(\omega_l/\omega_s)\cdot\int_tI_s(z,t)dt/\int_tI_l(z=0,t)dt$ vs. input pulse intensity (the same as vs. input pulse energy) both for co-propagating  ($z=L$) and counter-propagating ($z=0$) geometries.  A great increase of the conversion efficiency due to BW effect in the case of co-propagating waves is explicitly seen. Saturation at $I_l^0/I_{min}^p\approx7\cdot10^4 $ is due to depletion of fundamental radiation caused by conversion to Stokes radiation.
\begin{figure}[b]
\begin{center}
\includegraphics[width=.8\columnwidth]{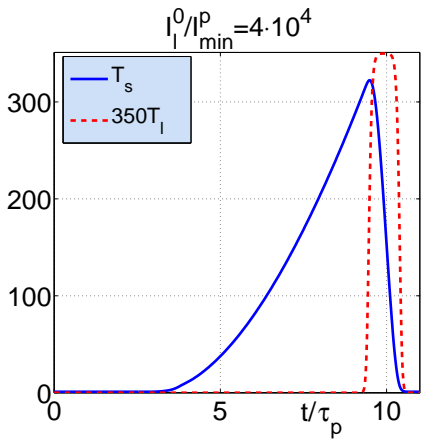}\\
\caption{\label{f6} Pulse shape of amplified Stokes radiation at the exit of crystal of length L = 1 cm,  $ T_s =\left| {{\mathcal{E} _s (L,t)}/{\mathcal{E} _s^0 }} \right|^2$  (solid line).  Pulse shape of the output fundamental pulse  $ T_l = \left| {{\mathcal{E} _l (L,t)}/{\mathcal{E} _l^0 }} \right|^2 $ is depicted by dashed line. Here, $ I_l ^ 0/I_ {min} ^ p = 4 \times10 ^ 4. $}
\end{center}
\end{figure}

\begin{figure}[t]
\begin{center}
\includegraphics[width=.8\columnwidth]{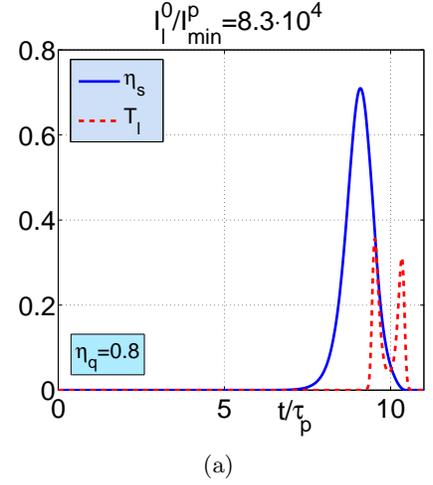}\\
(a)\\
\includegraphics[width=.8\columnwidth]{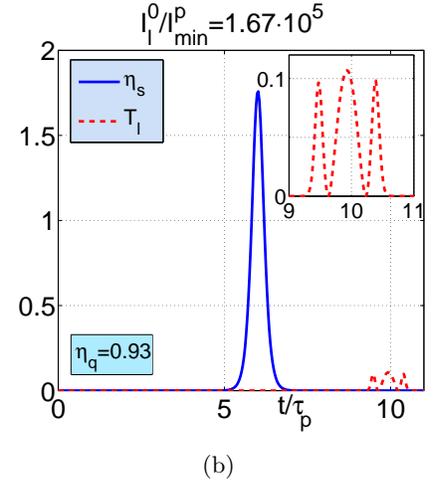}\\
(b)
\caption{\label{f7} Pulse shape of amplified Stokes radiation (solid line) and of transmitted fundamental radiation (dashed line) at the exit of crystal of L = 1 cm. $\eta_{s}=(\omega_l/\omega_s)\cdot \left| {{\mathcal{E} _s (L,t)}/{\mathcal{E} _l^0 }} \right|^2$  is conversion efficiency, $ T_l = \left| {{\mathcal{E} _l (L,t)}/{\mathcal{E} _l^0 }} \right|^2 $ is transmittance.  a. $ I_l ^ 0/I_ {min} ^ p = 8.3 \times10 ^ 4 $. b.  $ I_l ^ 0/I_ {min} ^ p = 1.67 \times10 ^ 5 $}
\end{center}
\end{figure}

Simulations also show that shape of the output Stokes and fundamental pulses  differ and vary significantly depending on the intensity of the input fundamental wave. Input seeding signal at Stokes frequency is assumed a weak CW broadband radiation.  Figure~\ref{f6} depicts  amplified output pulse  $ T_s =\left| {{\mathcal{E} _s (L,t)}/{\mathcal{E} _s^0 }} \right|^2$ of Stokes radiation for relatively small amplification of the signal and small depletion of the fundamental beam. Here, shape of the fundamental pulse remains unchanged. Pulse shape of amplified Stokes signal is different and determined by the fact that $v_s>v_l$ and Stokes pulse surpasses fundamental one.
In contrast, at  $ I_l ^ 0/I_ {min} ^ p = 8.3 \times10 ^ 4 $ and $ I_l ^ 0/I_ {min} ^ p = 1.67 \times10 ^ 5 $ (Figs.~\ref{f7}a,b),  conversion efficiency becomes significant ($ \eta_q = $ 0.8 and $ \eta_q = 0.93 $, respectively). Corresponding depletion of the output fundamental pulse  and changes in its shape is explicitly seen.
Note that in the case shown in Fig.~\ref{f7}(b), the output Stokes pulse significantly overtakes the pump pulse. In the latter case, major  conversion occurs in the middle of the medium, after which both pulses propagate almost without interacting and independently from each other. It is seen that output Stokes pulse narrows with increase of the input fundamental intensity.  Here, crystal length of 1cm corresponds to more than 1000 input pulse lengths ($T/t_p = L/L_p$ = 1357). Threshold intensity  $I^p_{min} = 6\cdot 10^6$ W/cm$^2$, which corresponds to 60 fs pulse of 5 $\mu$J focused to the spot of diameter D = 100 $\mu$m. Intensity of seeding Stokes signal was chosen   $I^0_s/I^0_l = 10^{-10}$.

Described  NLO propagation processes  are in  striking contrast with their counterparts in crystals where only phonons with positive group velocity exist \cite{ShB,Boy}. Such NLO properties  are also different from  those inherent to  phase-matched mixing of EM and acoustic waves for the cases where the latter have  energy flux and wave vector directed against  EM waves \cite{Bob}. Elaboration of the proposed concept  allows  to utilize  revealed extraordinary features for  creation of a family of unique photonic devices made of ordinary Raman crystals such as optical switches, filters, amplifiers  and cavity-free optical parametric oscillators. Proposed here concept is different from earlier proposed in \cite{Har} and does not require periodic poling of quadratic nonlinear susceptibility of crystals at the nanoscale as described in \cite{Kh} (and references therein).

\section{Conclusions}
 While the physics and applications of  NIMs are being explored world-wide at a rapid pace, current mainstream  focuses on fabrication of specially shaped nanostructures which enable negative optical magnetism. It is challenging task that relies on sophisticated methods of nanotechnology.  Engineering  a  strong  fast quadratic  NLO response by such mesoatoms also presents a  challenging goal not yet achieved. This paper proposes to mimic similar extraordinary backward-wave extraordinary nonlinear-optical propagation processes, however,  making use readily available Raman active crystals. Among the prospective application is a family of photonic devices with advanced functional properties. Basic underpinning idea is to replace one of the coupled  backward electromagnetic wave by optical phonons - elastic waves with  negative group velocity.Operation in short pulse regime is proposed to remove such a severe detrimental factor as  fast phonon damping. Significant decrease of the required minimum intensity of fundamental radiation compared with that in continuous-wave regime is shown down to that provided by commercial lasers. Unusual properties of the investigated processes are numerically simulated and the possibilities of pulse shape tailoring are predicted. There are many Raman active transparent crystals that support optical phonons with negative group velocity can be utilized, some of them, such as calcite, may offer further optimization of operational characteristics of the proposed photonic devices.

\begin{acknowledgments}
 This work was supported in parts by the Air Force Office of Scientific Research
(Contract  No FA950-12-1-298), by the  National Science Foundation (Grant No ECCS-1028353), by the
Academy of Finland and Nokia through the Center-of-
Excellence program, by the Presidium of the Russian
Academy of Sciences (Grant No 24-31), and by the
Russian Federal Program on Science, Education and In-
novation ( Grant No 14.A18.21.1942).
\end{acknowledgments}

\bibliographystyle{IEEEtran}

\end{document}